\documentclass[sn-mathphys,Numbered,iicol]{sn-jnl}

\usepackage{graphicx}%
\usepackage{multirow}%
\usepackage{amsmath,amssymb,amsfonts}%
\usepackage{amsthm}%
\usepackage{mathrsfs}%
\usepackage[title]{appendix}%
\usepackage{xcolor}%
\usepackage{textcomp}%
\usepackage{manyfoot}%
\usepackage{booktabs}%
\usepackage{algorithm}%
\usepackage{algorithmicx}%
\usepackage{algpseudocode}%
\usepackage{listings}%
\usepackage{mathptmx}
\usepackage{natbib}

\begin{document}

\title[Purcell's swimmer]{Dynamics of  Purcell-type microswimmers\linebreak with active-elastic joints}


\author*[1]{Anna Zigelman}\email{annar@technion.ac.il}

\author[1]{Gilad Ben Zvi}

\author[1]{Yizhar Or}\email{izi@technion.ac.il}


\affil[1]{Faculty of  Mechanical Engineering, Technion - Israel Institute of  Technology, Haifa 3200003, Israel}

%


\abstract{Purcell's planar three-link microswimmer is a classic model of swimming in low-Reynolds-number fluid, inspired by motion of flagellated microorganisms. Many works analyzed this model, assuming that the two joint angles are directly prescribed in phase-shifted periodic inputs. In this work, we study a more realistic scenario by considering an extension of this model which accounts for joints' elasticity and mechanical actuation of periodic torques, so that the joint angles are dynamically evolving. Numerical analysis of the swimmer's dynamics reveals multiplicity of periodic solutions, depending on parameters of the inputs - frequency and amplitude of excitation, joints' stiffness ratio, as well as joint's activation. We numerically study swimming direction reversal, as well as bifurcations, stability transitions, and symmetry breaking of the periodic solutions, which represent the effect of buckling instability observed in swimming microorganisms. The results demonstrate that this variant of Purcell's simple model displays rich nonlinear dynamic behavior with actuated-elastic joints. Similar results are also obtained when studying an extended model of a six-link microswimmer.}


%




\maketitle

\section{Introduction}\label{sec1}
Many living microorganisms such as sperm cells, nematodes and bacteria can swim in fluid environment by applying shape actuation of their elongated body or flagellated tail~\cite{Gray_1964,Gaffney_2011,Camalet_1999,Brokaw_1972,Machin_1958}. In recent years, these microscopic creatures have inspired the research and development of micro-robotic swimmers, an application of great importance e.g., in the field of biomedicine~\cite{Sitti_2015,Peyer_2013,Gao_2012}. In order to understand the basic locomotion mechanisms of these creatures and to improve the engineering design of such micro-robots, one needs to formulate basic mathematical models of microswimmer motion.

\begin{figure}
	\centering
	\includegraphics[width=0.4\textwidth]{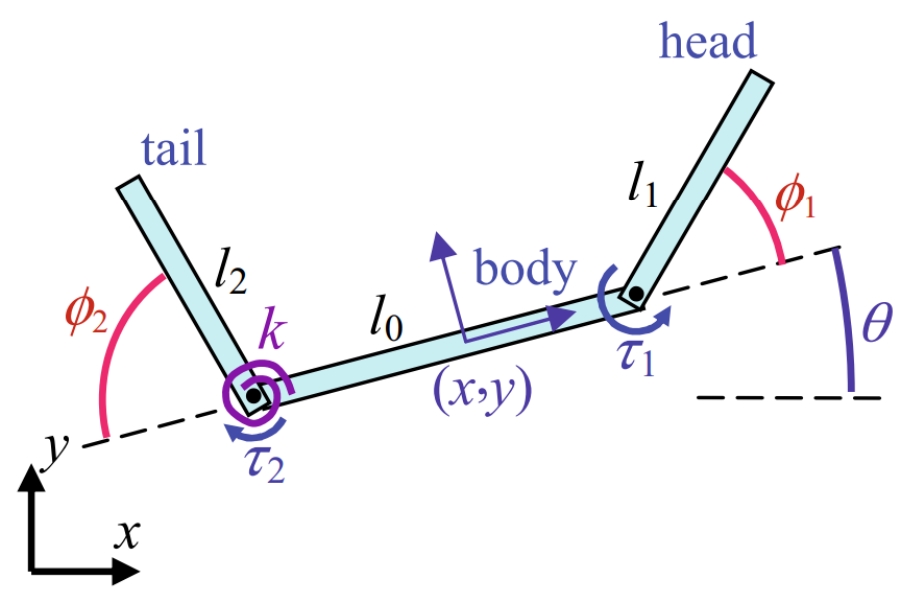}
	\caption{Purcell's three-link microswimmer model.} \label{fig1}
\end{figure}

A fundamental principle of microswimmers' motion is the fact that they move in the very low Reynolds number (Stokes flow), in which the dominant forces are generated by viscous drag resistance of the fluid while the influence of inertial effects is negligible, so that the motion is quasi-static~\cite{Happel_1983,Lauga_2009,Kim_2005}. Several basic models of microswimmers' motion were developed. For example, Taylor's first models included an analysis of a two-dimensional flexible sheet which moves as a propagating wave~\cite{Taylor_1951}, and based on this model he proposed a model of a swimmer with a ``helical tail'' whose shape is similar to a sperm cell. Later on, Purcell~\cite{Purcell_1977} proposed a model for a swimmer with a rotating helical tail and also presented a planar microswimmer model with three links connected by two rotational joints (Purcell swimmer), as shown in Fig.~\ref{fig1}. This swimmer was claimed in~\cite{Purcell_1977} to be the simplest mechanism that can move in a viscous flow by periodic actuation. The two inputs of periodic joint angles generate non-reversible time profile that enables net locomotion, as required by Purcell's ``scallop theorem''. Later, dynamic equations of motion for Purcell's swimmer were explicitly formulated under the assumption of slender links~\cite{becker_koehler_stone_2003}. Based on this model, an asymptotic analysis of the motion of Purcell's swimmer behavior under two types of periodic inputs (gaits) of joint angles was performed~\cite{Wiezel_2016}.  A macro-scale robotic implementation of Purcell's three-link swimmer was constructed and investigated~\cite{Gutman_2016}, for studying symmetry properties of various gaits. Other works utilized methods of geometric mechanics for analyzing the displacement under given gaits~\cite{Avron_2008,Hatton_2013}, whereas the works~\cite{Alouges_2019,Wiezel_2016A,Tam_2007,Nuevo-Gallardo,Wiezel_2021} utilized methods of optimal control for maximizing the swimmer's displacement or energy efficiency. All these studies were performed under the simplifying assumption that it is possible to kinematically control the swimmer's shape change, for example by directly dictating the joint angles. In the work~\cite{Or_2012}, the dynamics and stability of periodic solutions for joint angles of  Purcell swimmer were investigated under the assumption that the controlled inputs are periodic torques applied at the joints. However, in many practical situations, microswimmer's shape change is partially carried out passively and is also affected by an elastic response of the swimmer's structure. Several models have studied continuous bending deformation of elastic flagellum with distributed actuation in mathematical models~\cite{Spagnolie,Lauga_2007}, some of which were based on sperm cells motility~\cite{Dillon_2003,Gaffney_2011}. 
Such models revealed the existence of optimal ratio between stiffness and beating frequency that maximizes swimming performance. Another important effect that has been studied is \emph{dynamic buckling instability} in elastic-actuated swimming microorganisms and their theoretical models~\cite{Gadelha_2010,Lough_2023,Fily_2020,Son_2013,Gadelha_2020}. This phenomenon occurs when some parts of the swimmer's body have localized low stiffness (for example, a flexible hook at the tip of a rotating flagellum). In such case, increasing the actuation frequency may result in \emph{symmetry breaking bifurcation} where the symmetric beating of straight-line mean swimming loses stability and asymmetric configurations emerge, leading to curved swimming trajectories. The continuous models mentioned above typically involve partial differential equations, so that their analysis is relatively complicated. 
A simplified version of a model that takes into account the elastic effects of the body can be obtained by using rigid links connected by rotational joints with torsion springs. Hence, several such models have been proposed to describe the motion of flexible linked nano-swimmers actuated by an oscillating magnetic field~\cite{Gutman_2016_PRE,Gutman_2014_PRE,Harduf_2018,Alouges_2015}.

Passov and Or~\cite{Passov_2012} proposed a generalization of Purcell's three-link microswimmer, where one joint is passive elastic having a torsion spring, while the angle of the other joint is prescribed as a sinusoidal input. The model in~\cite{Passov_2012} displayed convergence of the passive angle to stable oscillations about zero mean (corresponding to the straightened load-free configuration of the torsion spring). In addition, this model exhibits an optimal frequency that maximizes the net displacement per cycle, which was also approximated in~\cite{Passov_2012} using asymptotic analysis. The work~\cite{Ramasamy_2021} considered the same model, and utilized geometric method and frequency-domain analysis in order to find optimal time-periodic input for maximizing either mean swimming speed or energy efficiency. A variant of the simple model in~\cite{Passov_2012} has been used in~\cite{Krishnamurthy_2017} for analyzing the motion of the swimming microorganism \emph{Schistosoma mansoni}, proving that it tunes the ratio between its structural stiffness and beating frequency in order to optimize performance.

Nevertheless, the above-mentioned models were not able to capture the dynamic buckling effect. A key reason for this shortcoming is that these models assumed kinematic input of prescribed joint angle at the active joint. In practice, a more realistic description involves mechanical actuation of torque applied at the joint, possibly combined with passive elasticity and/or feedback based on measurement of the actual joint angle. The goal of this work is to study such a variant of Purcell's microswimmer model with mechanical actuation, where both joint torques consist of time-periodic input signal and/or passive elastic terms. In such case, the two joint angles are evolving dynamically as a result of the swimmer-fluid interaction. This may lead to multiplicity of periodic solutions, with possible stability transitions, symmetry-breaking and bifurcations. This work presents numerical investigation of the influence of the swimmer's parameters on its dynamic behavior and characteristics of periodic solutions. We study the case in which both joints are active-elastic, as well as the case where only one joint is active. We show that when there is large ratio between joint stiffnesses, the swimmer's motion may undergo reversal in swimming direction, as well as symmetry-breaking bifurcations and stability transitions of periodic solutions, which represent the dynamic buckling effect. Lastly, we also present some examples of a higher-dimensional six-link model with elastic passive/active joints, showing a similar behavior.

\section{Problem statement}\label{sec2}
We now introduce Purcell's three-link microswimmer model~\cite{Purcell_1977} and the formulation of its dynamic equations of motion in viscous fluid~\cite{becker_koehler_stone_2003,Wiezel_2016,Gutman_2016}. The swimmer consists of three rigid links connected by two rotational joints, as shown in Fig.~\ref{fig1}. The coordinates of the robot are decomposed into body variables $\mathbf{q}=(x,y,\theta)^T$, which describe the position and orientation of the central link, and shape variables $\pmb{\Phi}=(\phi_1,\phi_2)^T$ which are the two relative angles at the joints. The vector of torques acting at the joints is denoted as $\pmb{\tau} = (\tau_1, \tau_2)^T$.

In low Reynolds number fluid, the viscous resistance force and torque acting on a rigid body in Stokes flow are linear in its translational and rotational velocities~\cite{Purcell_1977,Happel_1983}. For slender links, these expressions are given as~\cite{Wiezel_2016,becker_koehler_stone_2003,cox_1970}: 
\begin{equation}\label{E:res forces}
	\left.\begin{aligned}
		&\mathbf{f}_i=-c_t l_i(\mathbf{v}_i \cdot \mathbf{t}_i)\mathbf{t}_i-c_n l_i (\mathbf{v}_i \cdot \mathbf{n}_i)\mathbf{n}_i\\
		&m_i=-\frac{1}{12}c_n l_i^3\omega_i,
	\end{aligned}\right\}
\end{equation}
where $\mathbf{f}_i$ is the viscous drag force acting on link $i$, and $m_i$ is the hydrodynamic torque with respect to the link's center. $\mathbf{t}_i$ is a unit vector along the link's axial direction, $\mathbf{n}_i$ is a unit vector along transversal direction, $\mathbf{v}_i$ is the velocity vector of the link's center, and $\omega_i$ is its angular velocity along $\hat{\mathbf{z}}$ direction. 
For simplicity, we assume equal links lengths $l_0=l_1=l_2=l$. The resistance coefficients in~\eqref{E:res forces} for slender links are given as~\cite{cox_1970}:
\begin{equation}
	c_n=2c_t=\frac{4\pi\mu}{\log{(l/a)}},
\end{equation}
where $\mu$ is the fluid's viscosity and $a$ is the link's cross-section radius. Due to negligibility of inertial forces, the force and torque balance on the swimmer's links give rise to first-order equation of motion~\cite{Wiezel_2016}:
\begin{equation}\label{E:system1}	
	\dot{\mathbf{q}}=\mathbf{R}(\theta)\mathbf{G}(\pmb{\Phi})\dot{\pmb{\Phi}},
\end{equation} 
\begin{equation}\label{E:system2}	
	\dot{\pmb{\Phi}}=\mathbf{H}(\pmb{\Phi})\pmb{\tau},
\end{equation}
where $\mathbf{R}$ is the rotation matrix. The equation in~\eqref{E:system1} is the swimmer's equation of motion assuming kinematic input which directly prescribes the joint angles $\pmb{\Phi}(t)$, which was obtained in~\cite{becker_koehler_stone_2003,Wiezel_2016}. The equation in~\eqref{E:system2} is the relation between joint torques and the motion of the joint angles, which has been derived in~\cite{Gutman_2016,Or_2012,Passov_2012}.

The previous work~\cite{Passov_2012} assumed one passive elastic joint and one kinematically actuated joint with periodic input:
\begin{equation}
	\phi_1(t)=\varepsilon\sin(\omega t), \quad \tau_2=-k\phi_2.
\end{equation}
In this study, we assume that mechanical inputs are the joint torques, active and/or passive, where joint angles become dynamically evolving variables. Moreover, we assume that the swimmer is actuated by periodic torque input and parallel elasticity, so that the torques are expressed as
\begin{equation}\label{E:moment at joints}
	\tau_i=\varepsilon_i\sin(\omega t+\gamma_i)-k_i\phi_i=-k_i(\phi_i-\psi_i(t)), \quad i=1,2.
\end{equation}
This means that the joint torque contains periodic input connected in parallel to an elastic torsion spring having stiffness $k_i$. An equivalent description of the joint torque is a proportional feedback law which aims to make the angle $\phi_i$ track a desired periodic signal $\psi_i(t)$, as formulated in the last term of~\eqref{E:moment at joints}.

\section{Numerical integration results}\label{sec3}
In order to study the general behavior and to search for characteristic types of dynamic evolution, we perform numerical dynamic simulations on different combinations of the spring constants, frequencies, and actuation torque amplitudes. 

\begin{figure*}
	\centering
	\includegraphics[width=\textwidth]{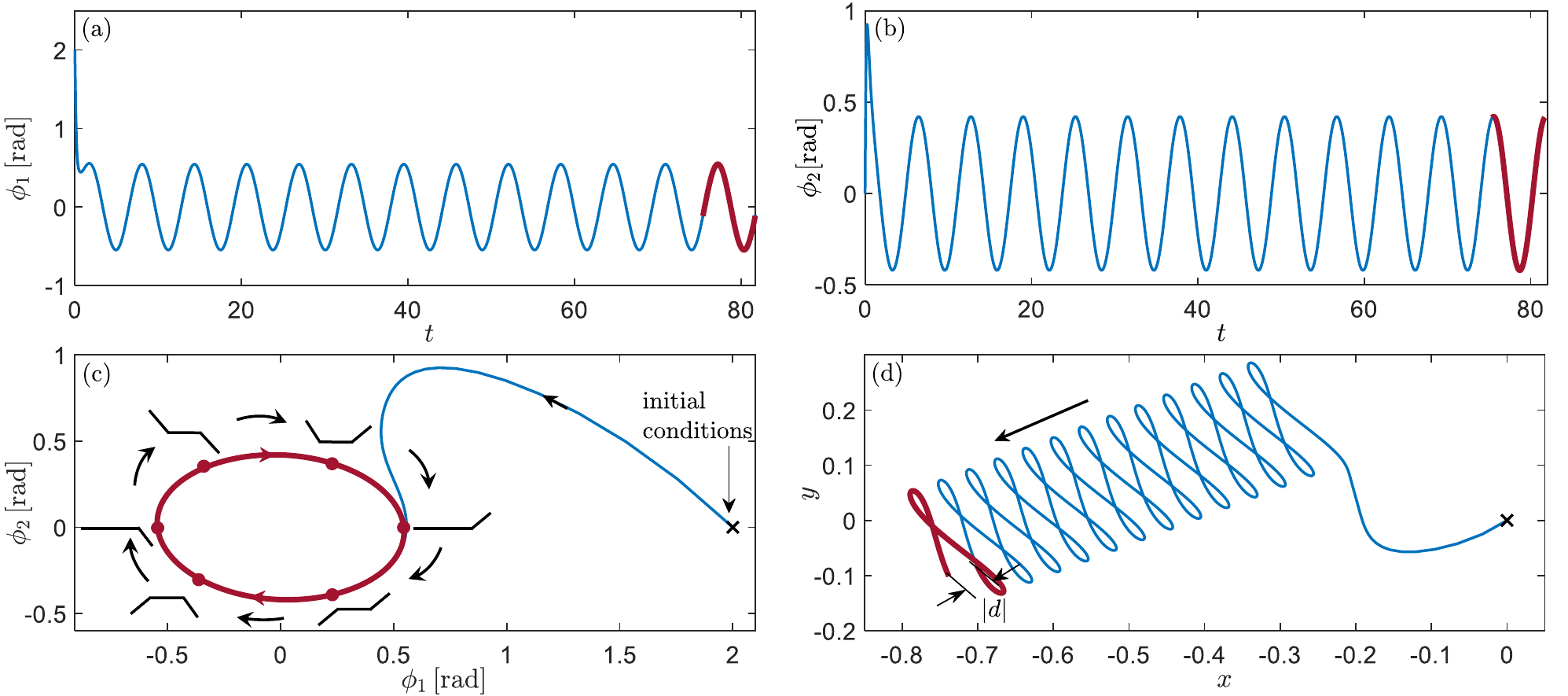}
	\caption{Simulations of {\bf{case 1}}, as given in~\eqref{E:case 1} - two identical active-elastic joints with periodic actuation torques. Joint angles (a) $\phi_1$ and (b) $\phi_2$ versus time $t$, as well as the swimmer's trajectory in (c) $\phi_1-\phi_2$ plane. The small swimmer's sketches illustrate the periodic shape changes during cycle. (d) The swimmer's trajectory in $x-y$ plane. The stable periodic symmetric solution is indicated with maroon color.} \label{fig2}
\end{figure*}

In order to render the problem in~\eqref{E:system1}--\eqref{E:moment at joints} dimensionless, we first define a characteristic scale of length $l$ and characteristic scale of torsional stiffness $k_c$ (units of moment). This induces a characteristic time scale~\cite{Passov_2012}, $t_c=c_t l^3/k_c$. Second, we scale all of the physical quantities by the characteristic scales to make them dimensionless:
\begin{equation}
	\begin{aligned}
	&\tilde{x}=\frac{x}{l}, \quad  \tilde{y}=\frac{y}{l}, \quad \tilde{t}=\frac{t}{t_c}, \quad \tilde{\omega}=\omega t_c, \quad \tilde{f}_i=\frac{f_i}{(c_t l^2/t_c)},\\	
	&\tilde{\tau}_i=\frac{\tau_i}{(c_t l^3/t_c)}, \quad \tilde{k}_i=\frac{k_i}{k_c}, \quad \tilde{\varepsilon}_i=\frac{\varepsilon_i}{k_c}.
	\end{aligned}
\end{equation}
For convenience, we hereafter remove the tilde ($\,\tilde{\,\,}\,$) symbol from all variables $\tilde{\tau}_i$, $\tilde{\varepsilon}_i$, $\tilde{k}_i$, and $\tilde{\omega}_i$, where we use the convention that they all represent scaled dimensionless quantities.

In order to perform dynamic simulation for the solution of the system in~\eqref{E:system1}, \eqref{E:system2}, \eqref{E:moment at joints} in different cases, we used the built-in ODE45 solver in Matlab.

In {\bf{case 1}}, we perform a simulation with torque-spring input, where both joints are actuated and have equal stiffnesses. The dimensionless input and parameter values are 
\begin{equation}\label{E:case 1}
	\begin{aligned}
	&\tau_1=A_1\sin{(\omega t)}-k_1\phi_1, \quad \tau_2=A_2\cos{(\omega t)}-k_2\phi_2,\\
	&k_1=k_2=1, \quad \omega=1, \quad A_1=A_2=0.5,
	\end{aligned}
\end{equation}
with the initial conditions $\phi_1(0)=2$ [rad] and \linebreak $\phi_2(0)=0$. In Fig.~\ref{fig2} we show the joint angles $\phi_1(t)$ and $\phi_2(t)$ versus time and the swimmer's trajectory in the $x-y$ plane and in the $\phi_1-\phi_2$ plane. It can be seen that after an initial transient, the solution converges to symmetric oscillations about zero mean angles,  $\bar{\phi}_i=0$, $i=1,2$, where the symmetry implies straight-line net swimming with net displacement per cycle denoted by $|d|$, as indicated in Fig.~\ref{fig2}(d).

\begin{figure*}
	\centering
	\includegraphics[width=\textwidth]{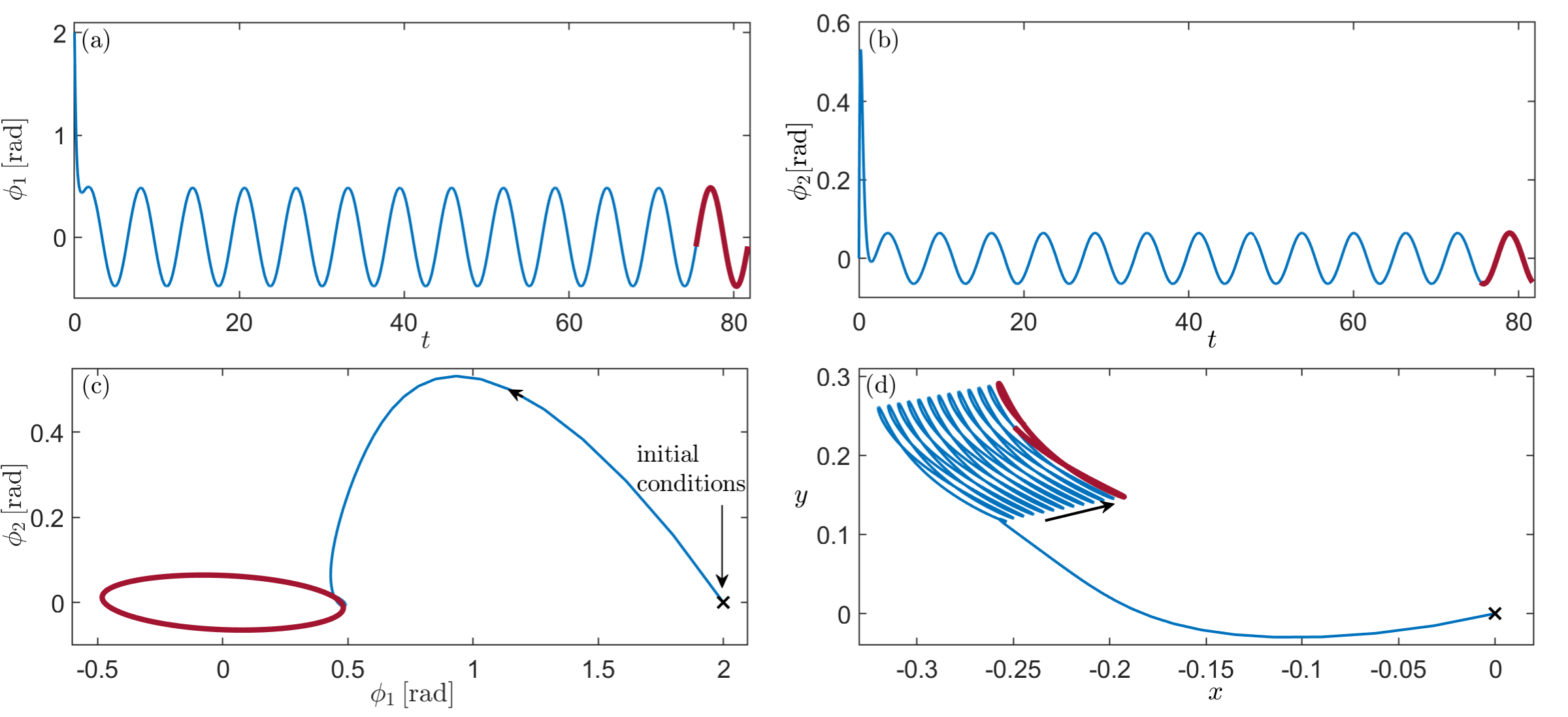}
	\caption{ Simulations of {\bf{case 2}}, as given in~\eqref{E:case 2} - actuated and passive elastic joints are with equal stiffnesses. Joint angles (a) $\phi_1$ and (b) $\phi_2$ versus time $t$, as well as the swimmer's trajectory in (c) $\phi_1-\phi_2$ and (d) $x-y$ plane. The stable periodic symmetric solution is indicated with maroon color.} \label{fig3A}
\end{figure*}

\begin{figure*}
	\centering
	\includegraphics[width=\textwidth]{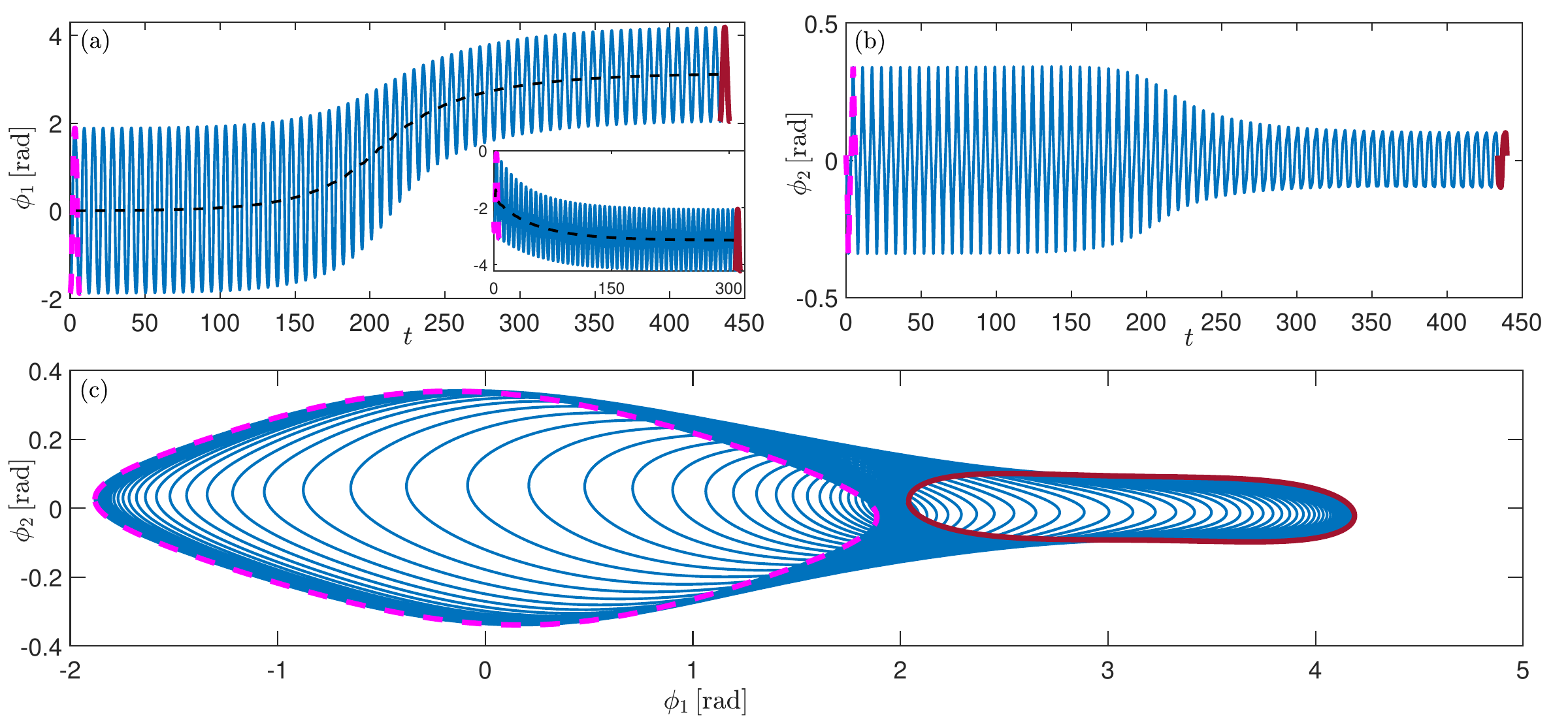}
	\caption{Simulations of {\bf{case 3}}, as given in~\eqref{E:case 3} - actuated joint with zero stiffness and passive elastic joint. Joint angles (a) $\phi_1$ and (b) $\phi_2$ versus time $t$, as well as (c) the swimmer's trajectory in  $\phi_1-\phi_2$ plane. The first period, which displays unstable symmetric oscillations about zero mean angles, is marked by a dashed magenta curve, and the stable periodic solution of oscillations about mean values of $(\bar{\phi}_1,\bar{\phi}_2)=(\pi,0)$ is indicated with solid maroon curve. The inset in (a) shows the convergence to the second solution with $(\bar{\phi}_1,\bar{\phi}_2)=(-\pi,0)$.} \label{fig3}
\end{figure*}

In {\bf{case 2}}, we perform a simulation with one active joint and one passive joint and with equal stiffnesses in both of them, namely
	\begin{equation}\label{E:case 2}
		\begin{aligned}
			&\tau_1=A_1\sin{(\omega t)}-k_1\phi_1, \quad \tau_2=-k_2\phi_2,\\
			&k_1=k_2=1, \quad \omega=1, \quad A_1=0.5, \quad A_2=0,
		\end{aligned}
\end{equation}
and with the initial conditions $\phi_1(0)=2$ [rad] and \linebreak $\phi_2(0)=0$. In Fig.~\ref{fig3A} we show the joint angles $\phi_1(t)$ and $\phi_2(t)$ versus time and the swimmer's trajectory in the $x-y$ plane and in the $\phi_1-\phi_2$ plane. Similarly to case 1 when the two joints were identical, also in this case (one joint is active and the other one is passive) it can be seen that after an initial transient, the solution converges to symmetric oscillations about zero mean angles,  $\bar{\phi}_i=0$, $i=1,2$, with straight-line net swimming. Moreover, when comparing between Figs.~\ref{fig2}(d) and \ref{fig3A}(d), it can be observed that the net displacement per cycle $|d|$ is much smaller in the case that one of the joints is passive, relative to the case that both joints are active.

In {\bf{case 3}}, we perform a simulation with one active joint and one passive joint and zero stiffness at the actuated joint, namely
\begin{equation}\label{E:case 3}
	\begin{aligned}
		&\tau_1=A_1\sin{(\omega t)}, \quad \tau_2=-k_2\phi_2,\\
		&k_1=0, \quad k_2=1, \quad \omega=1, \quad A_1=0.5, \quad A_2=0,
	\end{aligned}
\end{equation}
with the initial conditions $\phi_1(0)=-1.88$ [rad] and $\phi_2(0)=0.0189$ [rad] for the main figure in Fig.~\ref{fig3} (and $\phi_1(0)=-2.5$ [rad] and $\phi_2(0)=-2$ [rad] for the inset). 
In Fig.~\ref{fig3} we show the joint angles $\phi_1(t)$ and $\phi_2(t)$ versus time and the swimmer's trajectory in $\phi_1-\phi_2$ plane. It can be seen that the swimmer's joint angles initially perform symmetric oscillations about mean zero values $\bar{\phi}_i=0$. However, the oscillations of $\phi_1(t)$ eventually drift from mean zero, and converge to oscillations about $(\bar{\phi}_1, \bar{\phi}_2) = (\pi,0)$. Thus, we conclude that the symmetric periodic solution with \linebreak $\bar{\phi}_i=0$ is unstable and there is a convergence to stable periodic solution with mean $(\bar{\phi}_1,\bar{\phi}_2)=(\pi,0)$, where two links are folded on each other. Note that the latter oscillations also result in straight-line net motion due to symmetry about the swimmer's folded configuration. Note that a configuration with $\phi_1=\pi$ where two links are folded on top of each other is unphysical and unreachable in a planar setup due to intersection of the links, and also irrelevant for undulating microorganisms. Nevertheless, such configuration may still be possible for an articulated robot which is designed such that the links are moving in parallel planes. Additionally, we shall see below that this case serves as a mathematical limit of cases where the stiffness of the active joint is very low, which lead to asymmetric periodic solutions.

\begin{figure}[ht!]
	\centering
	\includegraphics[width=0.5\textwidth]{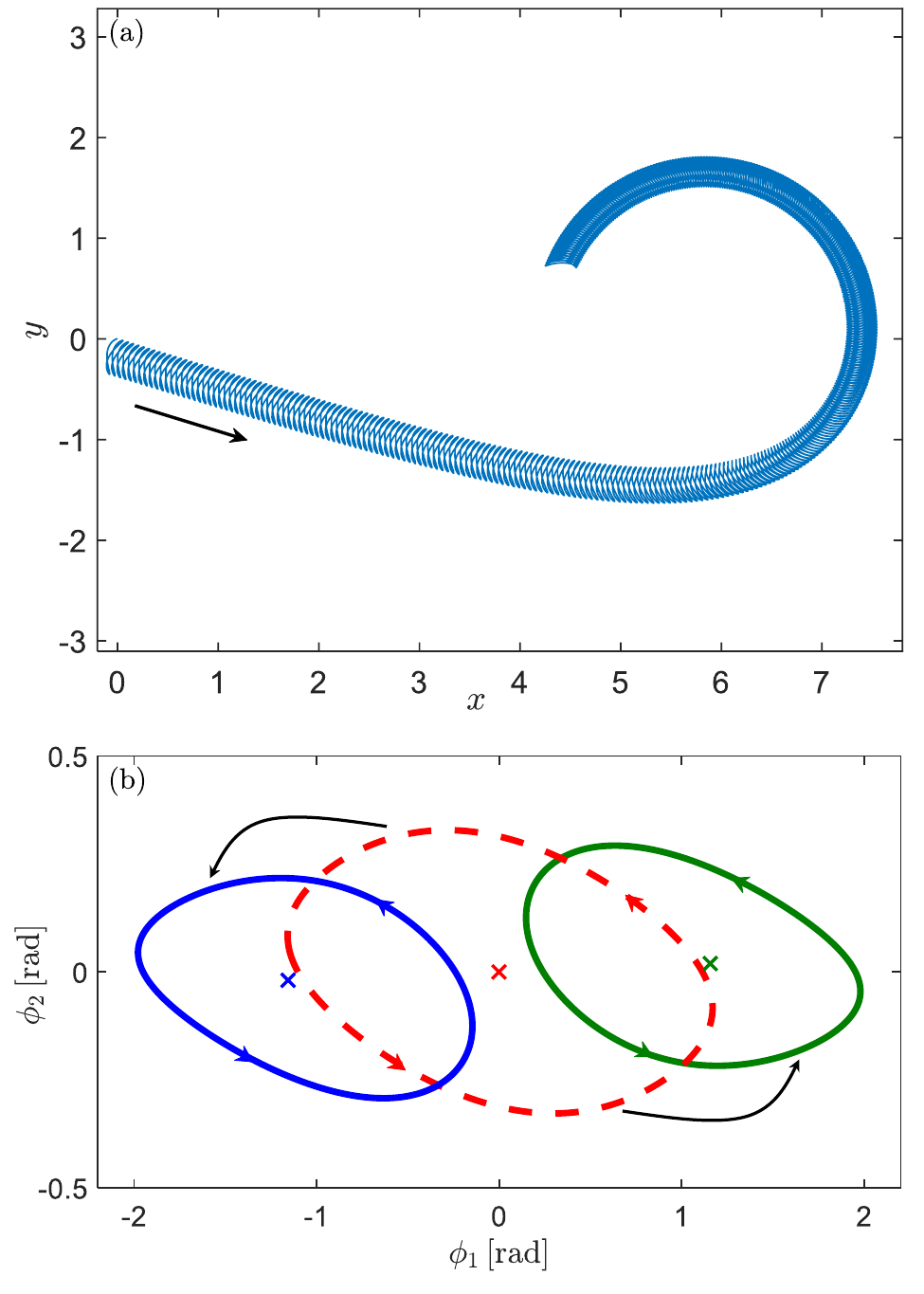}
	\caption{Simulations of {\bf{case 4}}, as given in~\eqref{E:case 4} - actuated joint with low stiffness and passive elastic joint. The swimmer's trajectory in (a) $x-y$ plane and (b) $\phi_1-\phi_2$ plane. The red dashed curve represents unstable symmetric solution, while the green and blue solid curves represent stable asymmetric solutions. The colored 'x' markers denote the mean values $(\bar{\phi_1},\bar{\phi_2})$ corresponding to the different periodic solutions. Setting the initial conditions close to the values along the dashed red curve, the solution will converge either to the green or to the blue curve.} \label{fig4}
\end{figure}

Considering the periodic solution of symmetric oscillations about straightened configuration $\bar{\phi}_i=0$, we observed above that it is stable for sufficiently large stiffness $k_1$ (cases 1 and 2), whereas it is unstable for zero stiffness (case 3). Thus, it is reasonable to assume that there exists an intermediate value of $k_1$ for which a stability transition occurs. Hence, in {\bf{case 4}}, we perform a simulation with a small value of $k_1$. Specifically, in this case we use the following input,
\begin{equation}\label{E:case 4}
	\begin{aligned}
		&\tau_1=A_1\sin{(\omega t)}-k_1\phi_1, \quad \tau_2=-k_2\phi_2,\\
		&k_1=0.01, \quad k_2=1, \quad \omega=2, \quad A_1=0.5, \quad A_2=0,
	\end{aligned}
\end{equation} 
with initial conditions $(\phi_1(0),\phi_2(0))\approx (-1.154,0.048)$ for the unstable symmetric solution, and $(\phi_1(0),\phi_2(0))=(0.166,0.0702)$ and $(\phi_1(0),\phi_2(0))=(-1.975,0.028)$ for the two asymmetric solutions (all angles are given in [rad]). In Fig.~\ref{fig4} we show the swimmer's trajectory in $x-y$ plane and in $\phi_1-\phi_2$ plane. It is possible to see that the swimmer starts the motion with oscillations about the straightened configuration $\bar{\phi}_i=0$, which generates motion along straight line for some cycles. Then the solution eventually diverges away and converges to asymmetric oscillations about nonzero mean values, $\bar{\phi}_i\neq 0$. Due to asymmetry of the oscillations, they result in swimming along a circular arc, with nonzero net rotation, as shown in Fig.~\ref{fig4}. Note that in this case, unstable symmetric periodic solutions with $\bar{\phi}_i= 0$ co-exist. Interestingly, for parameter values as given in equation~\eqref{E:case 4}, this behavior is obtained only for specific range of actuation frequencies, $1.16< \omega < 8.77$. For higher or lower frequencies, one obtains only stable symmetric oscillations about $\bar{\phi}_i =0$, as in case 2. This motivates the analysis of periodic solutions, as presented next. 

\section{Analysis of periodic solutions (Poincar\'{e} map)}\label{sec4}
After we have seen various possible behaviors of the swimmer in response to different torque inputs and discussed situations which appear to be classified as stable or unstable periodic solutions of symmetric or asymmetric oscillations, as well as multiplicity of periodic solutions for given configurations, in this section, we shall investigate periodic solutions and their stability in a rigorous way. In particular, we scan the possible parametric ranges in order to find and classify periodic solutions.

In order to investigate periodic solutions we use Poincar\'{e} map~\cite{Guckenheimer}, which is found numerically. Let us define the system's state as $\mathbf{z}=(\phi_1,\phi_2)$ and define $\mathbf{z}_k=\mathbf{z}(t=kt_p)$, where $t_p=2\pi/\omega$  and $k=1,2,3,\ldots$. Poincar\'{e} map is then defined as $\mathbf{F}(\mathbf{z}_k)=\mathbf{z}_{k+1}$, which induces a discrete-time-invariant dynamical system for $\mathbf{z}_k$. Initial conditions $\mathbf{z}(0)=\mathbf{z}^*$ which lead to a periodic solution of~\eqref{E:system2}, $\mathbf{z}(t)=\mathbf{z}(t+t_p)$, are reflected by a fixed point of the Poincar\'{e} map, namely $\mathbf{F}(\mathbf{z}^*)=\mathbf{z}^*$. After a periodic solution corresponding to fixed point $\mathbf{z}^*$ is found, it is possible to use the eigenvalues of the Jacobian matrix of the Poincar\'{e} map $\mathbf{J}=d\mathbf{F}/d\mathbf{z}|_{\mathbf{z}=\mathbf{z}^*}$ in order to determine whether the solution is locally stable or unstable. Specifically, if all eigenvalues of the Jacobian satisfy $|\lambda_i(\mathbf{J})|<1$, then it is possible to conclude that the periodic solution is locally asymptotically stable~\cite{Guckenheimer}. However, if there exists at least one eigenvalue satisfying $|\lambda_i|>1$, then the solution is unstable. In summary, it is possible to find periodic solutions by numerical search for solutions of $\mathbf{F}(\mathbf{z}^*)=\mathbf{z}^*$. Moreover, this method also allows to find multiplicity of periodic solutions and to analyze their stability.

\begin{figure*}
	\centering
	\includegraphics[width=\textwidth]{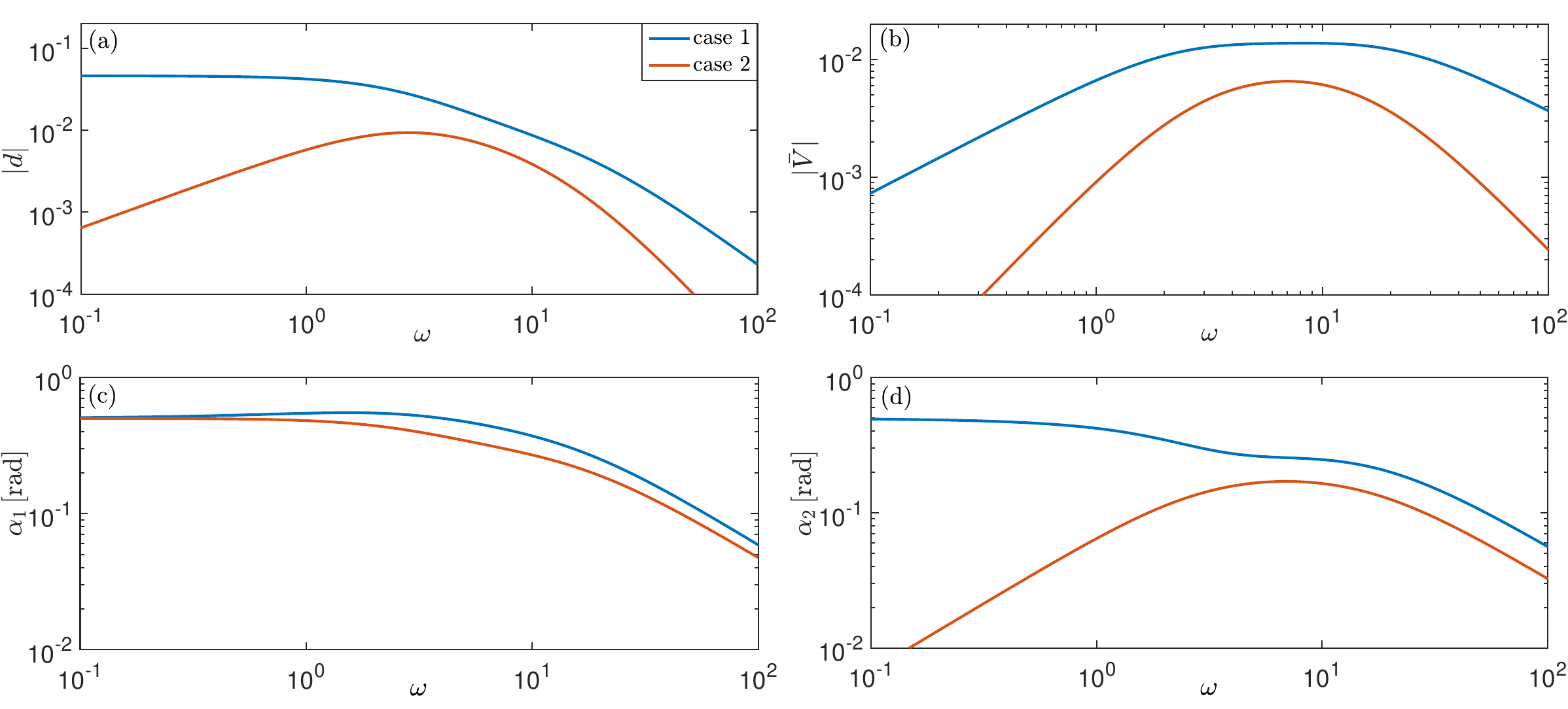}
	\caption{Properties of the symmetric periodic solutions as a function of actuation frequency $\omega$, where for the blue curve the parameters correspond to {\bf{case 1}} and are given in~\eqref{E:case 1} except that $\omega$ varies in $[0,100]$ and for the orange curve the parameters correspond to {\bf{case 2}} and are given in~\eqref{E:case 2} except that $\omega$ varies in $[0,100]$: (a) Displacement per cycle $|d|$, (b) mean speed $|\bar{V}|$, (c) oscillation amplitude of $\phi_1(t)$ denoted by $\alpha_1$, and (d)  oscillation amplitude of $\phi_2(t)$ denoted by $\alpha_2$. All of the panels are in log-log scale.} \label{fig5}
\end{figure*}

For symmetric solutions only, let us define the mean value $\bar{\theta}$ about which the angle $\theta(t)$ oscillates after initial transient response followed by convergence to a periodic solution, as the mean value of $\theta(t)$ over a period, namely
\begin{equation}
	\bar{\theta}=\int_{t_0}^{t_0+t_p}\theta(t)dt.
\end{equation}
Note that for symmetric solutions $\bar{\theta}$ is a constant that does not depend on the choice of $t_0$, after settling to steady-state periodic oscillations. Let us define signed net displacement per cycle, which we denote by $d$, as follows
\begin{equation}
	d=\left(\begin{aligned}&x(t_0+t_p)-x(t_0)\\
		&y(t_0+t_p)-y(t_0)\end{aligned}\right)\cdot \left(\begin{aligned}&\cos{(\bar{\theta})}\\
		&\sin{(\bar{\theta})}\end{aligned}\right),
\end{equation}
where $\cdot$ denotes the scalar product.
Then the signed mean speed per cycle which we denote by $\bar{V}$ may be defined as
\begin{equation}
	\bar{V}=\frac{d}{t_p}.
\end{equation}

By varying system's dimensionless parameters $(A_1,A_2,k_1,k_2,\omega)$, we are able to find multiple periodic solutions and analyze their stability and bifurcations. First, we consider the cases of equal stiffnesses at the two active joints or at one active and one passive joint (see cases 1 and 2 which are given in~\eqref{E:case 1} and~\eqref{E:case 2}), and vary the actuation frequency $\omega$. 
In Fig.~\ref{fig5} we show the absolute value of the net displacement per cycle $|d|$, the absolute value of the mean speed $|\bar{V}|$, and oscillation amplitudes of $\phi_i(t)$ in steady state, which are denoted by $\alpha_i$, versus frequency for two sets of parameters, as given in~\eqref{E:case 1} and~\eqref{E:case 2}. Note that for both these sets of input parameters there exists a unique solution of symmetric oscillations about mean zero angles, $\bar{\phi}_i = 0$, which is stable. We did not obtain multiple or unstable periodic solutions for any dimensionless frequency in the range $\omega\in[0,100]$. Moreover, it is possible to see that for case 1 where both joints are active, namely $A_1=A_2=0.5$, the net displacement $|d|$ grows monotonically with the actuation frequency $\omega$, but there exists optimal frequency that gives maximum mean speed $\bar{V}$ (Fig.~\ref{fig5}). However, for case 2 with one active joint and one passive joint, namely $A_1=0.5$ and $A_2=0$, there exists an optimal frequency which allows to obtain maximal net displacement per cycle $|d|$ and a different optimal frequency, that allows to obtain maximal mean speed $|\bar{V}|$, see Fig.~\ref{fig5}. This observation is different from the results in~\cite{Passov_2012}, where the actuated joint involved direct kinematic input of the angle $\phi_1(t)$ as time-periodic signal, which resulted in optimal frequency for maximizing $|d|$, but not for maximizing $|\bar{V}|$.

An additional observation is that in cases 1 and 2 (of equal stiffnesses) the direction of motion of the swimmer is constant and not reversed as a function of $\omega$. However, generally a reversal in the direction of motion when varying $\omega$ can occur. Let us show this phenomenon in the following examples, which we denote by {\bf{case}} $\bf{5}$ with the following parameters,
\begin{equation}\label{E:case 5A}
	\begin{aligned}
		&\tau_1=A_1\sin{(\omega t)}-k_1\phi_1, \quad \tau_2=A_2\cos{(\omega t)}-k_2\phi_2,\\
		&k_1=0.1 \quad \text{or} \quad k_1=0.01, \quad k_2=1, \quad A_1=A_2=0.5, 
	\end{aligned}
\end{equation}
and varying frequency, $\omega\in[0,100]$. In this case, both joints are active but the ratio between their stiffnesses is large.

\begin{figure*}
	\centering
	\includegraphics[width=\textwidth]{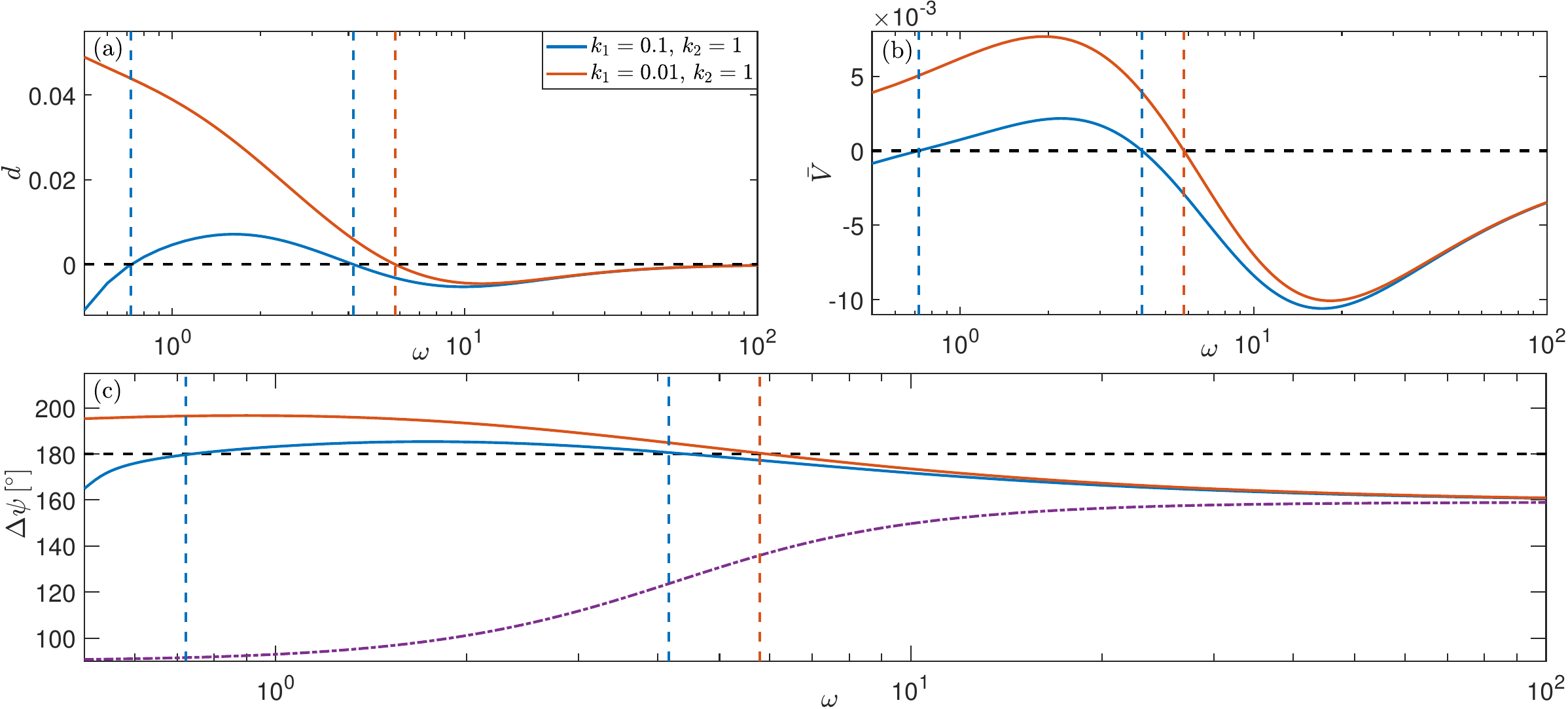}
	\caption{Properties of the symmetric periodic solutions in {\bf{case 5}} as a function of actuation frequency $\omega$, where the parameters are given in~\eqref{E:case 5A}, for the blue curve $k_1=0.1$, and for the orange curve $k_1=0.01$: (a) Signed displacement per cycle $d$, (b) mean speed $\bar{V}$, and (c) phase difference $\Delta \psi$ between steady-state oscillations of the two joint angles $\phi_1(t)$ and $\phi_2(t)$. All panels are in log-linear scale. The horizontal black lines in (a) and (b) denote zero displacement or velocity, respectively, and in (c) phase difference of $180^{\circ}$. The vertical dashed lines denote sign reversal of the signed distance. The dashed-dotted purple curve in panel (c) denotes $\Delta \psi$ for {\bf{case 1}} with $k_1=k_2=1$, which does not cross $180^{\circ}$.} \label{fig12}
\end{figure*}

In Fig.~\ref{fig12} we show the signed net displacement per cycle $d$, the mean speed $\bar{V}$, and the phase difference $\Delta \psi$ between steady-state oscillations of the two joint angles $\phi_1(t)$ and $\phi_2(t)$, for the two values of $k_1$ in case 5 given in~\eqref{E:case 5A}. It can be seen that if $k_1$ is sufficiently small $k_1\leq 0.1$, then there exists a frequency $\omega$ for which the swimmer reverses its direction of motion and starts to move in the opposite direction. In order to explain this direction reversal, we now show that it is tightly related to the phase difference between the oscillations of the two joint angles. Fig.~\ref{fig12}(c) shows the phase difference $\Delta \psi$ as a function of frequency, for the two values of $k_1$ in case 5, as well as for case 1 with two equal stiffnesses $k_1=k_2=1$. In case 1, the phase difference of the joint angles changes monotonically between $90^{\circ}$ and $180^{\circ}$, so that $\phi_1(t)$ lags behind $\phi_2(t)$, resulting in net motion towards joint 2, i.e. $d<0$. However, for case 5 with $k_2=1$ and $k_1=0.1$ or $k_1=0.01$, the phase difference crosses $180^{\circ}$, which means interchange in roles of leading and lagging joint angles, implying reversal in direction and in sign of $d$. Moreover, note that for $k_1=0.1$, the net displacement attains global maximum between two sign reversals. On the other hand, for $k_1=0.01$, the net displacement has a single reversal and does not have a local maximum.

\begin{figure*}
	\centering
	\includegraphics[width=\textwidth]{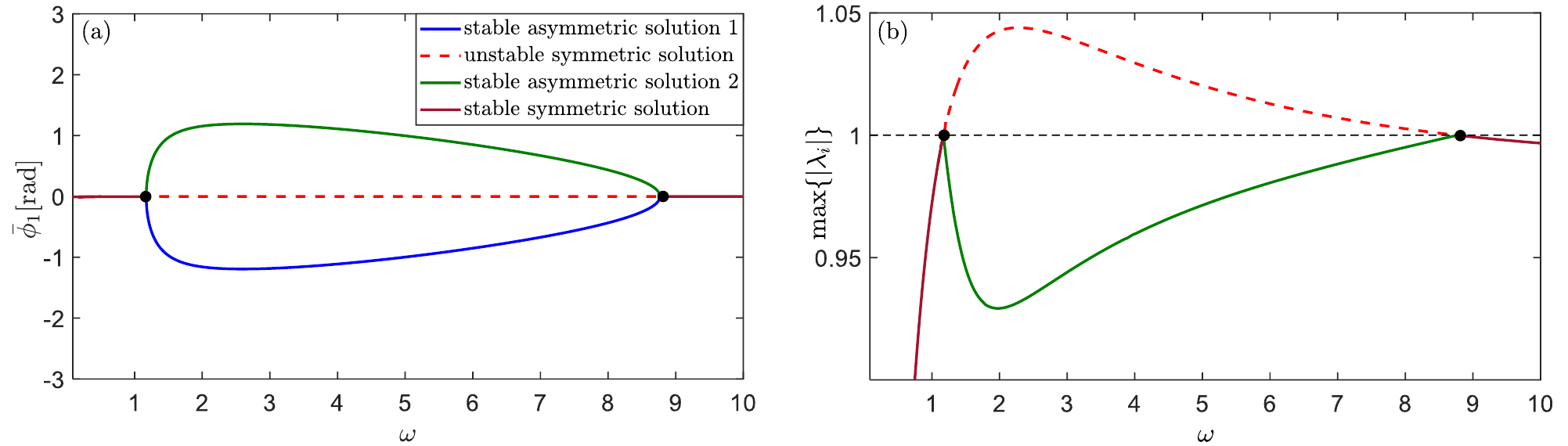}
	\caption{Properties of the symmetric periodic solutions in {\bf{case 4}} as a function of actuation frequency $\omega$, on log-linear scale, where the parameters are given in~\eqref{E:case 4}: (a) Mean value of $\phi_1$ and (b) the maximum of absolute values of eigenvalues $\lambda_i$. The 
		$\{\text{solid, dashed}\}$ curves represent the $\{\text{stable, unstable}\}$ symmetric solutions, respectively. The green and blue curves represent stable asymmetric solutions, and the filled black circles indicate the bifurcation points. The dashed black line in (b) represents a constant equal to 1.} \label{fig6}
\end{figure*}

\begin{figure*}
	\centering
	\includegraphics[width=\textwidth]{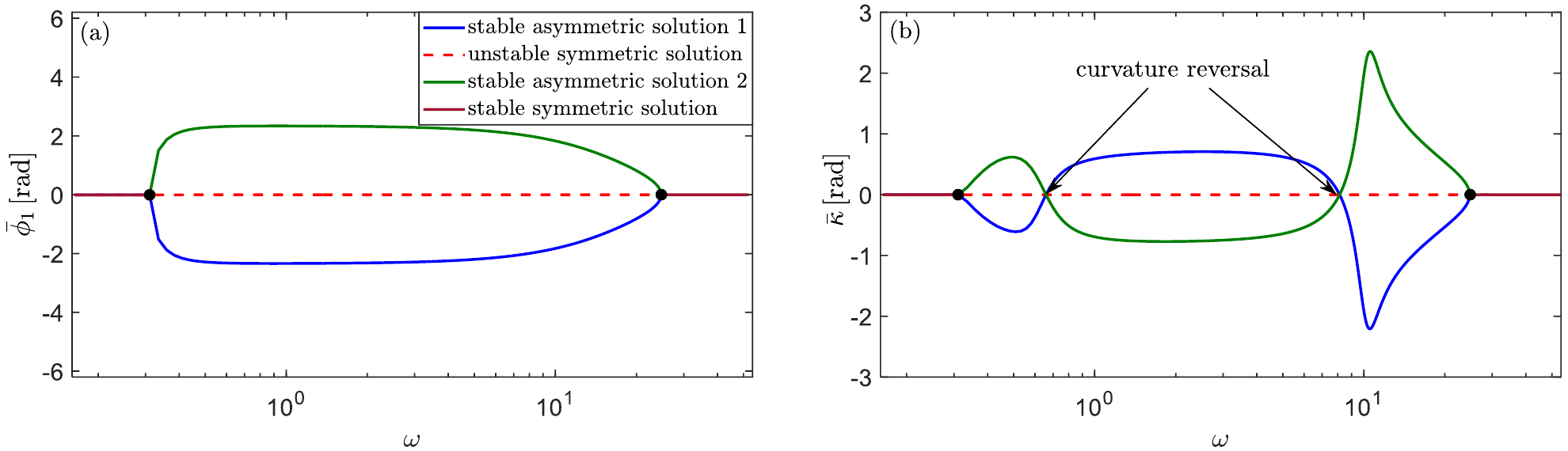}
	\caption{Properties of the symmetric periodic solutions in {\bf{case 6}} as a function of actuation frequency $\omega$, on log-linear scale, where the parameters are given in~\eqref{E:case 5}: (a) Mean value of $\phi_1$ and (b) the trajectory net curvature $\bar{\kappa}=\Delta\theta/d$. The 
		$\{\text{solid, dashed}\}$ curves represent the $\{\text{stable, unstable}\}$ symmetric solutions, respectively. The green and blue curves represent stable asymmetric solutions, and the filled black circles indicate the bifurcation points.} \label{fig7}
\end{figure*}

Next, we revisit case 4 where one joint is passive and the actuated joint has low dimensionless stiffness, with parameter values given in~\eqref{E:case 4}. Since our previous simulations show different behavior depending on dimensionless actuation frequency (see Fig.~\ref{fig4}), we now vary $\omega$ within the range $[0.085,10]$ and study properties of the multiple periodic solutions. Fig.~\ref{fig6} shows plots of the mean angle $\bar{\phi}_1$, and maximal eigenvalue magnitude $\max{|\lambda_i|}$ as a function of $\omega$, for the multiple periodic solutions. It can be seen that there exist three different frequency ranges: in the range of small and large frequencies only stable symmetric periodic solution with $\bar{\phi}_i=0$ exists, whereas in the intermediate range the symmetric solution becomes unstable and a pair of stable asymmetric periodic solutions emerge. At the two transition frequencies we obtain supercritical pitchfork bifurcation~\cite{Strogatz}. Note that this phenomenon is similar to the \emph{symmetry-breaking bifurcation} observed in swimming microorganisms and their continuous theoretical models~\cite{Gaffney_2011,Gadelha_2010,Lough_2023,Fily_2020,Son_2013,Gadelha_2020}. As already observed in these works, this effect is tightly related to \emph{dynamic buckling instability}, which occurs when one joint (the active one) has a very low stiffness.

We now analyze the influence of actuation frequency for {\bf{case 6}} where the dimensionless stiffness of the active joint is even smaller, namely, the input parameters are given by
\begin{equation}\label{E:case 5}
	\begin{aligned}
		&\tau_1=A_1\sin{(\omega t)}-k_1\phi_1, \quad \tau_2=-k_2\phi_2,\\
		&k_1=0.001, \,\, k_2=1, \,\, A_1=0.3, \,\, A_2=0, \,\, \omega\in[0,55].
	\end{aligned}
\end{equation}  

\begin{figure*}[ht!]
	\centering
	\includegraphics[width=\textwidth]{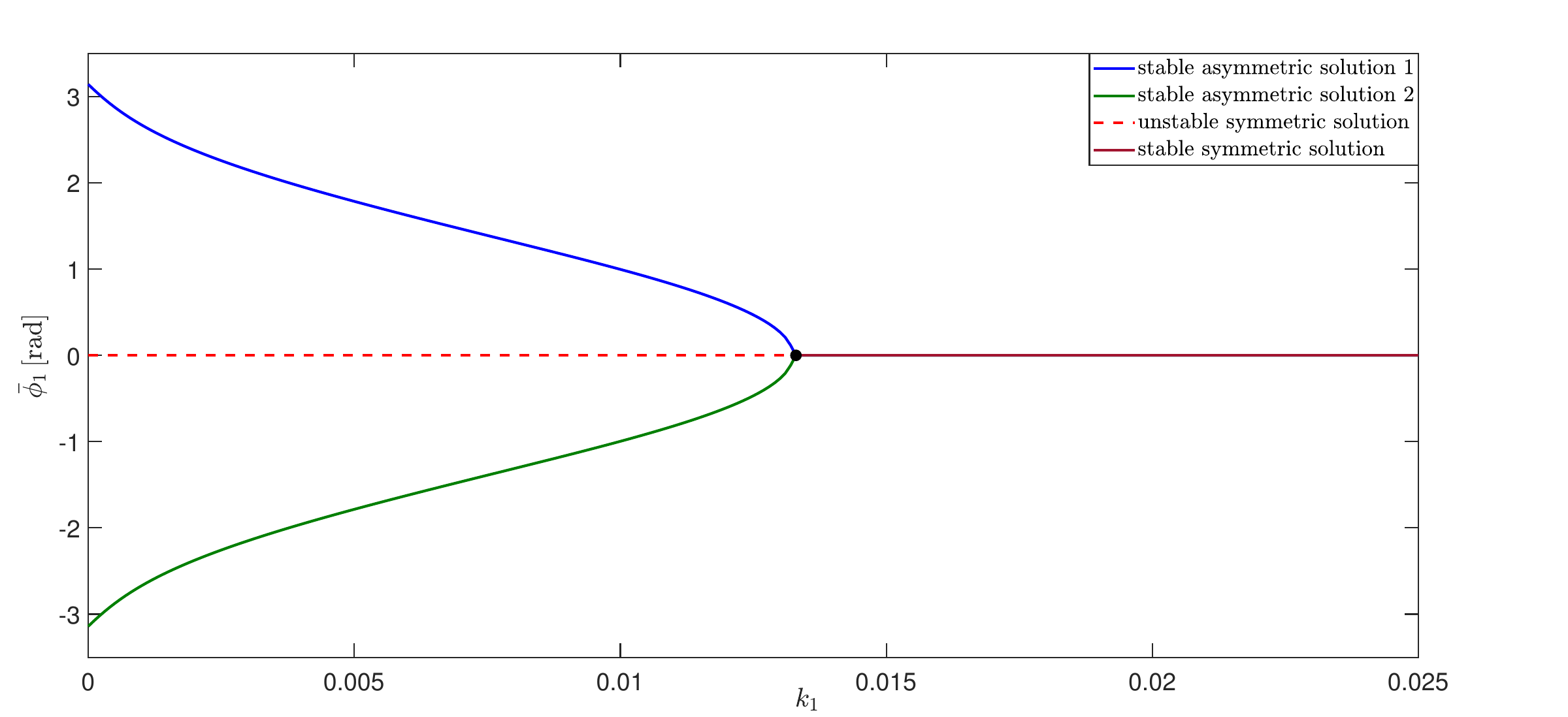}
	\caption{Mean value of the joint angle $\phi_1$ versus actuated joint's dimensionless stiffness $k_1$ for the input given in~\eqref{E:case 7}. The $\{\text{solid, dashed}\}$ curves represent the $\{\text{stable, unstable}\}$ symmetric solutions, respectively. The green and blue curves represent stable asymmetric solutions, and the filled black circle indicates the bifurcation point.} \label{fig8}
\end{figure*}

Fig.~\ref{fig7}(a) plots the mean value $\bar{\phi}_1$ for the multiple periodic solutions as a function of actuation frequency $\omega$. The system displays similar bifurcation structure as previous case shown in Fig.~\ref{fig6}.  Fig.~\ref{fig7}(b) plots the net curvature of periodic solutions, defined as $\bar{\kappa}= \Delta \theta /d$, as a function of frequency $\omega$, where  $\Delta \theta=\theta(t+t_p)-\theta(t)$ is the net swimmer's rotation per cycle. Interestingly, it can be seen that the curvature crosses zero and reverses its sign while varying $\omega$. This means that steering the swimmer's trajectory following curved paths in plane can be made possible under symmetric input, by simply varying the actuation frequency.

\begin{figure*}[ht!]
	\centering
	\includegraphics[width=\textwidth]{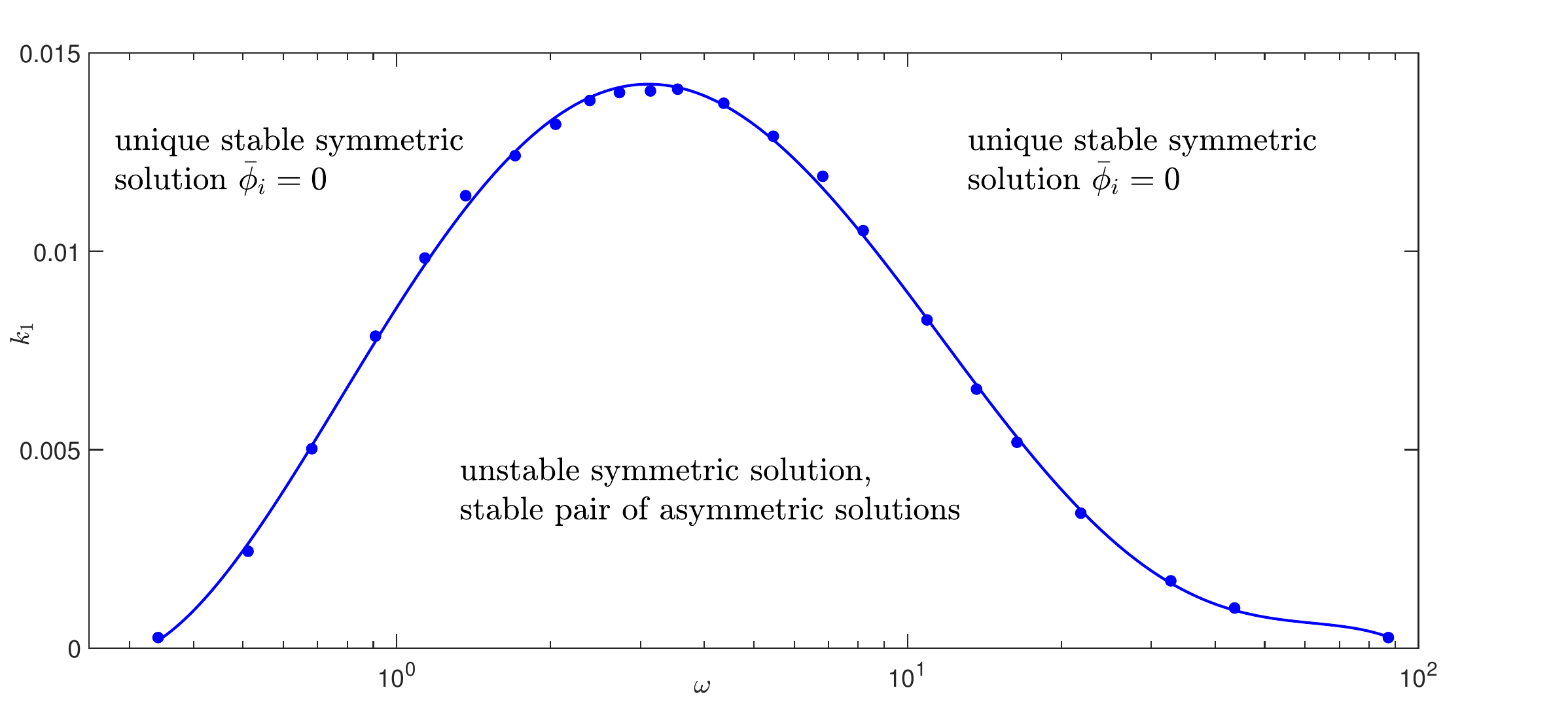}
	\caption{Stability limit curve of the symmetric periodic solution in $k_1-\omega$ plane, where $A_1=0.5$ and $A_2=0$ are kept fixed. The blue dots are the bifurcation points obtained by numerical simulation, and the solid blue line (obtained in Matlab by least square fitting to a 6-order polynomial) is to guide the eyes.} \label{fig9}
\end{figure*}

Finally, we consider the effect of varying the dimensionless stiffness $k_1$ of the active joint, while $k_2$ and the actuation frequency $\omega$ are kept fixed. More specifically, we consider input and parameter values given by
\begin{equation}\label{E:case 7}
	\begin{aligned}
		&\tau_1=A_1\sin{(\omega t)}-k_1\phi_1, \quad \tau_2=-k_2\phi_2,\\
		&k_2=1, \,\, \omega=5, \,\, A_1=0.5, \,\, A_2=0, \,\, k_1\in[0,0.025].
	\end{aligned}
\end{equation}  
Fig.~\ref{fig8} shows the mean joint angle $\bar{\phi}_1$ in different periodic solutions, as a function of active joint's stiffness $k_1$. It can be seen that for large $k_1$, in the same order of $k_2$, there exists only stable symmetric solution with $\bar{\phi}_i = 0$. When $k_1$ is decreased below a critical value, a pitchfork bifurcation occurs where a pair of asymmetric periodic solutions emerge, while the symmetric solution becomes unstable, as demonstrated in Figs.~\ref{fig4}, \ref{fig6}, and~\ref{fig7}. In the limit of $k_1 \to 0$, the mean values of active joint's angle in the asymmetric solution branches tend to $\bar{\phi}_1 \to \pm \pi$, a folded configuration, in agreement with the simulation results of case 2, shown in Fig.~\ref{fig3}.

From the stability analysis obtained by varying $(k_1,\omega)$ while keeping $A_1=0.5$ and $A_2=0$, we have numerically produced a graph of stability limit curve of the symmetric periodic solution $\bar{\phi}_i=0$ in $k_1-\omega$ plane, which is shown in Fig.~\ref{fig9}. It is possible to conclude that in the region above the limiting curve, single symmetric solution is stable, whereas in the region below the limiting curve the symmetric solution is unstable and there exists a pair of stable asymmetric solutions.

\section{Six-link swimmer}
\begin{figure}[b!]
	\centering
	\includegraphics[width=0.48\textwidth]{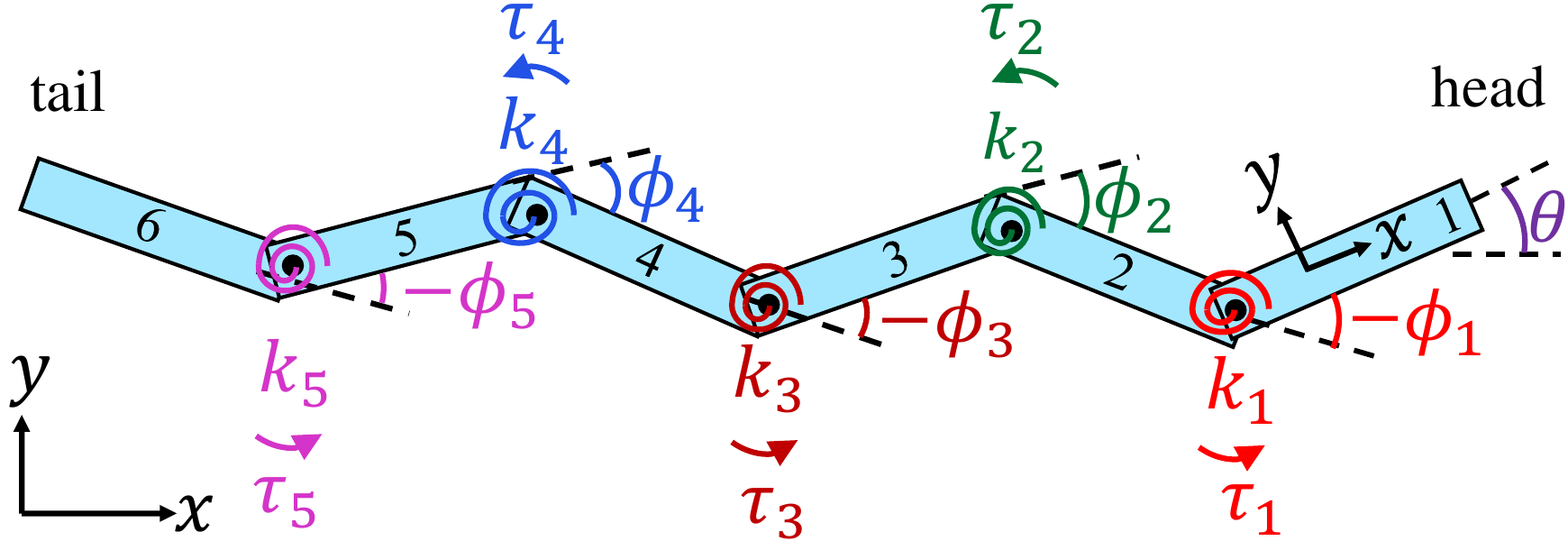}
	\caption{The six-link swimmer. Joint angles $\phi_i$, $i=1,\ldots,5$, denote the relative angle between links $i+1$ and $i$, so that they are positive in counter-clockwise direction.} \label{fig_new_sketch}
\end{figure}

We now consider a six-link microswimmer, which is a natural generalization of Purcell's swimmer discussed above. The six-link swimmer consists of six rigid links connected by five rotational joints, as shown in Fig.~\ref{fig_new_sketch}. In this case, the coordinates are decomposed into body variables $\mathbf{q}=(x,y,\theta)^T$, which describe the position and orientation of the first link (referred hereafter as ``head''), and shape variables $\pmb{\Phi}=(\phi_1,\phi_2,\phi_3,\phi_4,\phi_5)^T$ which are the five angles at the joints between the corresponding neighboring links. The vector of torques acting at the joints is denoted as $\pmb{\tau} = (\tau_1, \tau_2,\tau_3,\tau_4,\tau_5)^T$. All of the assumptions that we used for Purcell's swimmer throughout this study, including low Reynolds number, remain valid for the six-link swimmer model. Thus, we use the natural generalization of the problem in~\eqref{E:system1}-\eqref{E:system2} to 8 variables given in $\mathbf{q}$ and $\pmb{\Phi}$ corresponding to the six-link swimmer. The dynamic equations of motion for the six-link microswimmer model with active-elastic joints were formulated using the general scheme presented in~\cite{Wiezel_2016}, which is also similar to the elastic multi-link model studied in~\cite{Harduf_2018}.

\begin{figure*}[h!]
	\centering
	\includegraphics[width=\textwidth]{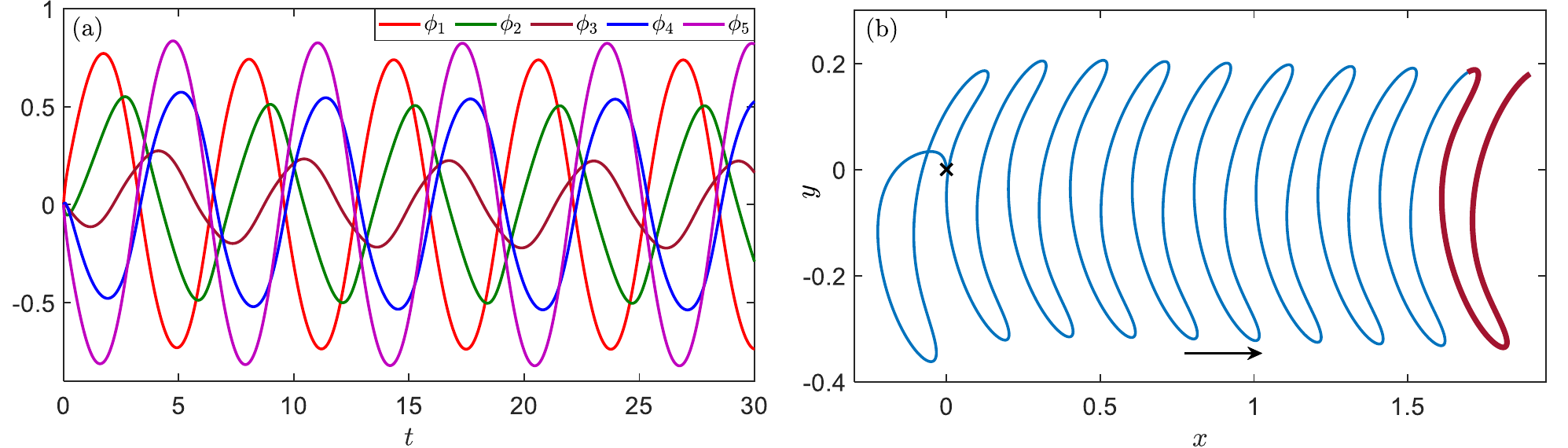}
	\caption{Simulations of six-link swimmer in {\bf{case 7}}, as given in~\eqref{E:case 8} - all 5 joints are actuated with equal actuation amplitudes, equal stiffnesses, and phase difference of $\psi=\pi/5$ between the adjacent actuated angles. (a) Joint angles $\phi_i$, $i=1,2,\ldots,5$, versus time $t$ and (b) the swimmer's trajectory in the $x-y$ plane. The stable periodic solution is indicated by maroon color in panel (b).} \label{fig10}
\end{figure*}

\begin{figure*}[h!]
	\centering
	\includegraphics[width=\textwidth]{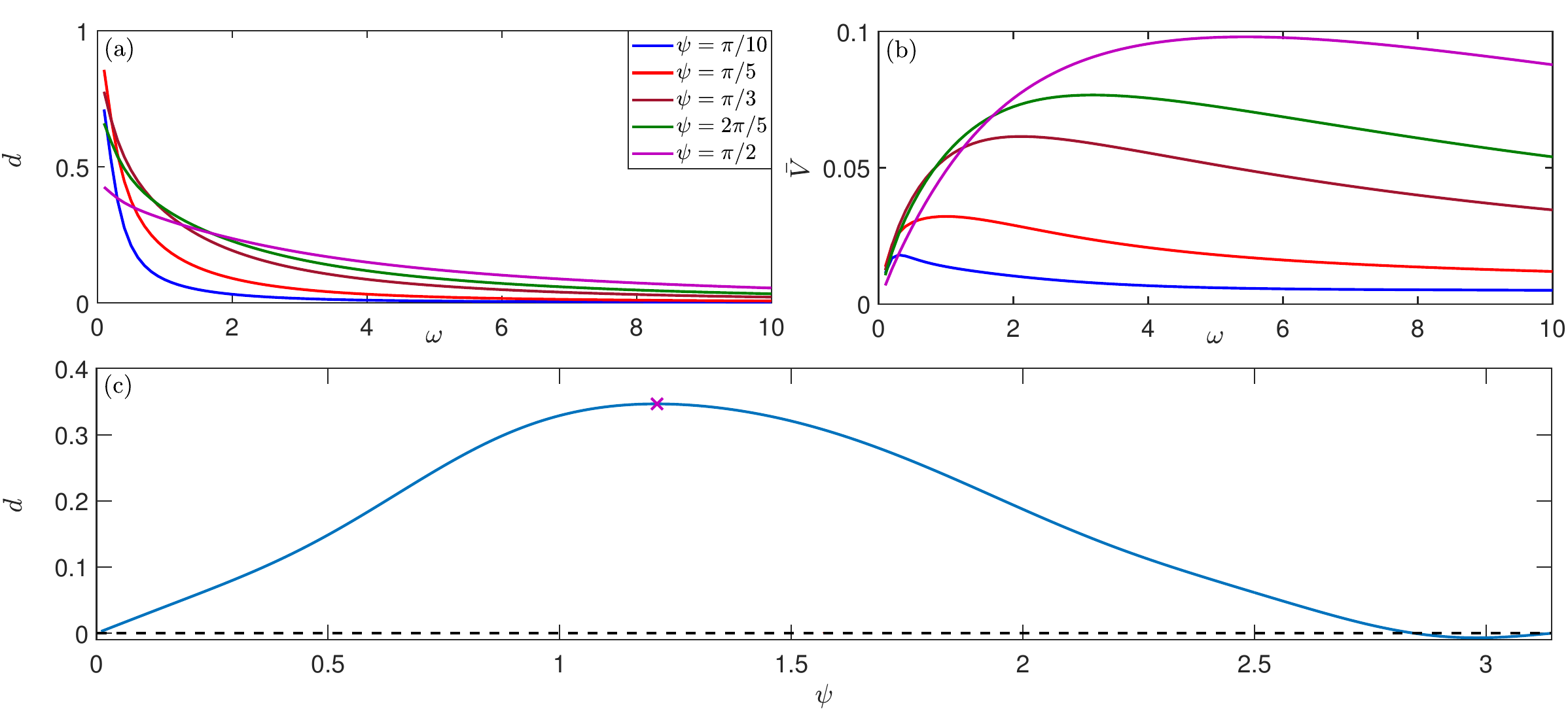}
	\caption{Properties of the symmetric periodic solutions in {\bf{case 7}}, as given in~\eqref{E:case 8}. (a) Displacement per cycle $d$ and (b) mean speed $\bar{V}$ versus $\omega$, where the colors of the curves represent the results for various phase differences $\psi$, and (c) displacement per cycle $d$ versus $\psi$ for $\omega=1$. The maximum and zero displacement in panel (c) are obtained for $\psi\approx0.4\pi$ and $\psi\approx0.9\pi$, respectively. The black dashed line in panel (c) denotes zero displacement.} \label{fig11}
\end{figure*}

We simulated the dynamics of the six-link swimmer in several cases. In the first case, which we denote by {\bf{case 7}}, the input parameters are:
\begin{equation}\label{E:case 8}
	\begin{aligned}
		&\tau_i=A_i \sin{(\omega t-(i-1)\psi)}-k_i\phi_i,	\quad \omega=1,\\
		&A_i=1, \quad k_i=1, \quad i=1,\ldots,5, \quad \psi=\pi/5,
	\end{aligned}
\end{equation}
with the initial conditions $\phi_i(0)=0$, $i=1,\dots,5$. In this case, all joints have equal stiffnesses, and all are actuated by periodic torque inputs with constant phase lag $\psi$ between consecutive joints.
In Fig.~\ref{fig10} we show the joint angles $\phi_i(t)$, $i=1,\dots,5$ versus time and swimmer's trajectory in the $x-y$ plane. It can be seen, that similarly to cases 1 and 2 in the three-link swimmer (see Figs.~\ref{fig2} and~\ref{fig3A}), after an initial transient the solution converges to symmetric oscillations about zero mean angles, $\bar{\phi}_i=0$, $i=1,\dots,5$, where the symmetry implies straight-line swimming with net displacement per cycle $|d|$. Moreover, it can be observed that the steady-state oscillation amplitude of $\phi_i(t)$ is minimal in the middle of the swimmer (for $i=3$) and increases towards both ends, the head and tail joints. 

\begin{figure*}[h!]
	\centering
	\includegraphics[width=\textwidth]{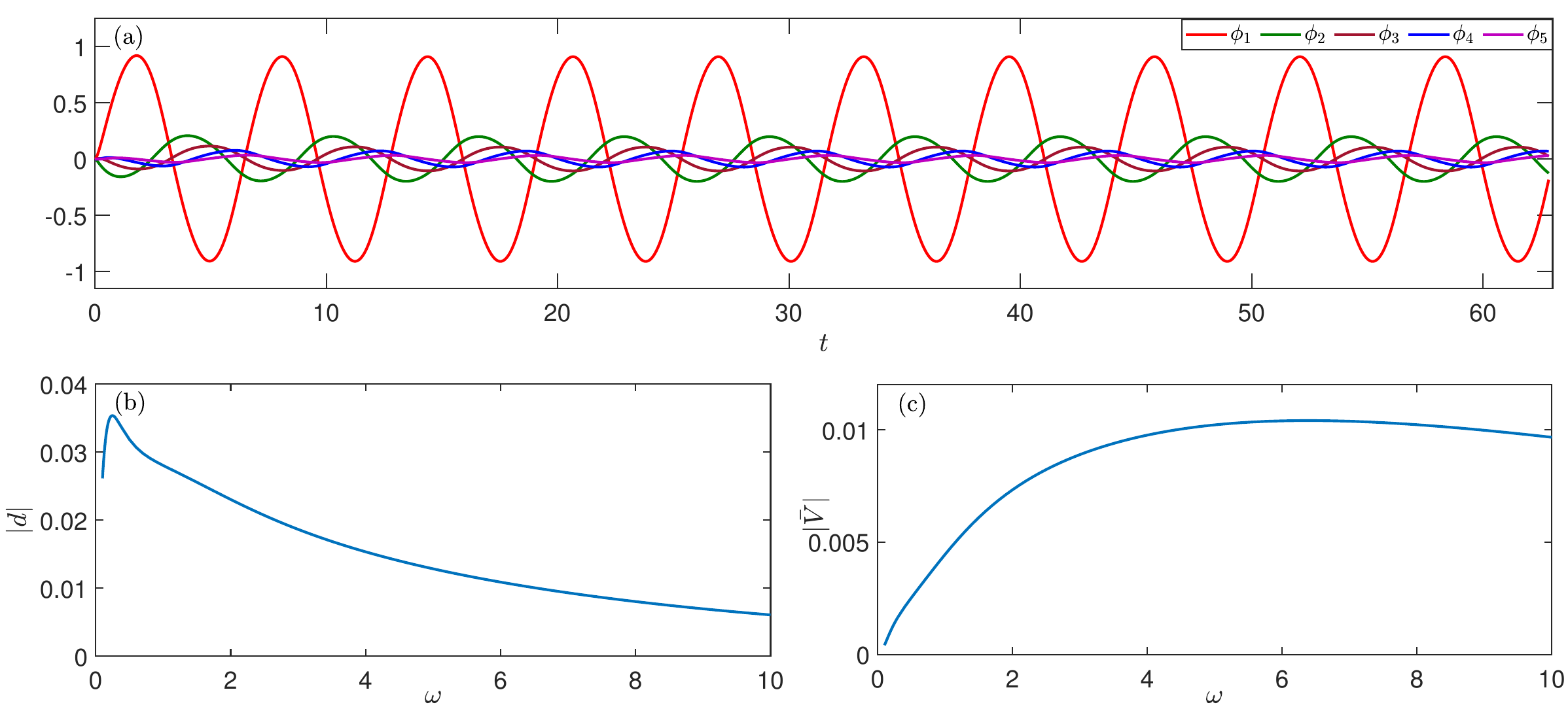}
	\caption{Simulations of six-link swimmer in {\bf{case 8}}, as given in~\eqref{E:case 9} - only one joint is actuated and the rest of the joints are passive ($A_1=1$, $A_i=0$ for $i>1$), where the stiffnesses of all joints are equal ($k_i=1$). (a) Joint angles $\phi_i$, $i=1,2,\ldots,5$, versus time $t$ for $\omega=1$, (b) displacement per cycle $|d|$ and (b) mean speed $|\bar{V}|$ versus $\omega$.} \label{fig13}
\end{figure*}

In Fig.~\ref{fig11}(a), (b) we show the net displacement per cycle $d$ and the mean speed per cycle $\bar{V}$ in {\bf{case 7}} versus the frequency $\omega$ for various phase differences $\psi$. In Fig.~\ref{fig11}(a), (b) it can be seen that for all phase differences $\psi$ the net displacement decreases monotonically with increase of $\omega$, so that there does not exist an optimal frequency for which maximal displacement is obtained, whereas there exists an optimal frequency for which maximal mean speed is obtained. This result is similar to Purcell's swimmer with two actuated joints (see Fig.~\ref{fig5}-case 1 of the three-link swimmer).

An additional interesting observation is that $d$ and $\bar{V}$ are not monotonic functions of the phase difference $\psi$. This fact is further visualized in Fig.~\ref{fig11}(c), which shows $d$ as a function of phase difference $\psi$ for fixed frequency $\omega=1$.
In particular, it can be seen that the maximal displacement (for $\omega=1$) is attained for $\psi\approx 0.385\pi$ and displacement crosses $d=0$ and changes sign for $\psi\approx0.9055\pi$. Note that based on our results for Purcell's swimmer, we are already familiar with the fact that the displacement may change sign, as e.g., was shown in Fig.~\ref{fig12}.

Next, we examine the case where all joints are passive except the first one, which is actuated. Specifically, the input parameters in {\bf{case 8}} are:
\begin{equation}\label{E:case 9}
	\begin{aligned}
		&\tau_i=A_i \sin{(\omega t)}-k_i\phi_i,	\quad \omega=1,\\
		&A_1=1, \quad 
		A_2=A_3=A_4=A_5=0, \\
		&k_i=1, \quad \text{for $i=1,\ldots,5$},
	\end{aligned}
\end{equation}
with the initial conditions $\phi_i(0)=0$, $i=1,\dots,5$. In Fig.~\ref{fig13}(a) we show the joint angles $\phi_i(t)$, $i=1,\dots,5$ versus time and in Fig.~\ref{fig13}(b), (c) we show the net displacement $|d|$ per cycle and mean speed $|\bar{V}|$ versus $\omega$. It can be seen, that similarly to cases 1, 2, and 7 (see Figs.~\ref{fig2}, \ref{fig3A}, and~\ref{fig10}), after an initial transient the solution converges to symmetric oscillations about zero mean angles, $\bar{\phi}_i=0$, $i=1,\dots,5$. Moreover, note that the oscillation amplitudes of the joints are decreasing from head to tail. Furthermore, in this case there exist two different optimal frequencies for which maximal displacement or maximal mean speed are obtained. This result is similar to Purcell's swimmer with one actuated joint (see Fig.~\ref{fig5}-case 2 of the three-link swimmer).

\begin{figure*}[h!]
	\centering
	\includegraphics[width=\textwidth]{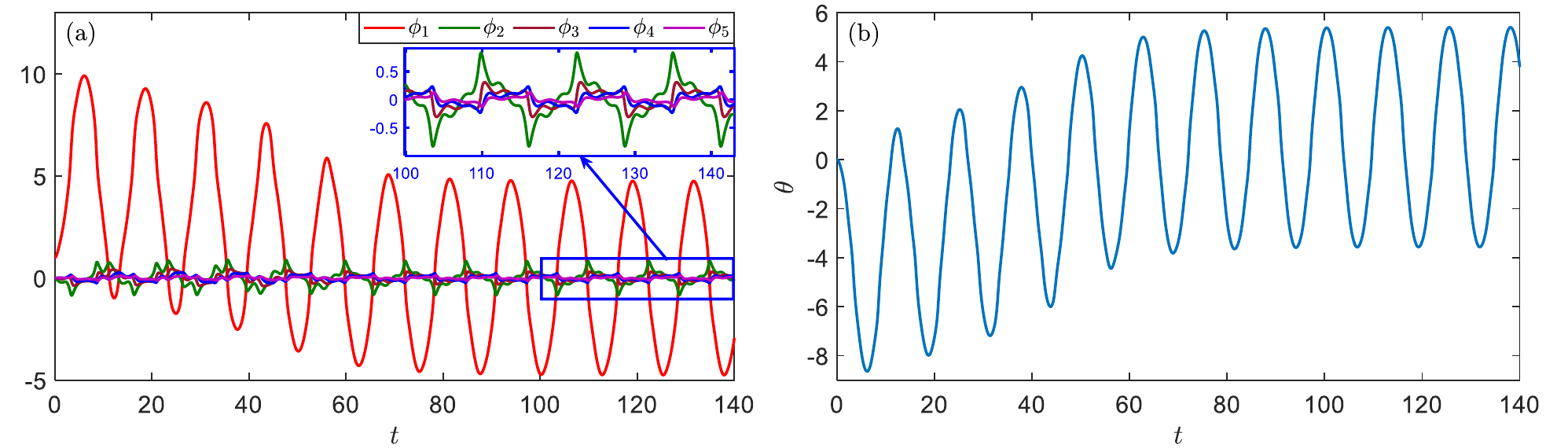}
	\caption{Simulations of six-link swimmer in {\bf{case 9}}, as given in~\eqref{E:case 10} - only one joint is actuated and the rest of the joints are passive ($A_1=1$, $A_i=0$ for $i>1$), where $k_1=0.01$ is much smaller than the rest of the stiffnesses ($k_i=1$ for $i>1$) for $\omega=0.5$. (a) Joint angles $\phi_i$, $i=1,2,\ldots,5$, versus time $t$, where in the inset we show zoom-in of the same plot (for joint angles $\phi_i$, $i=2,\ldots,5$). (b) The head angle $\theta$ versus $t$. The steady-state periodic solution is symmetric about $\bar{\phi}_i=0$ and $\theta=const.$, leading to straight-line swimming.} \label{fig14}
\end{figure*}

\begin{figure*}[h!]
	\centering
	\includegraphics[width=\textwidth]{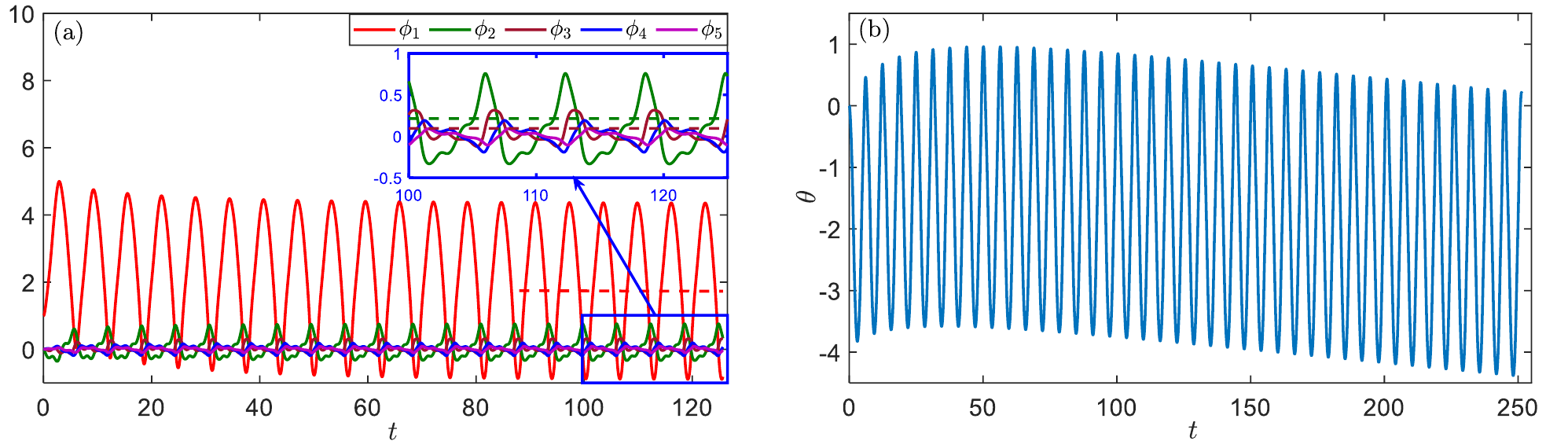}
	\caption{Simulations of six-link swimmer in {\bf{case 9}}, as given in~\eqref{E:case 10} - only one joint is actuated and the rest of the joints are passive ($A_1=1$, $A_i=0$ for $i>1$), where $k_1=0.01$ is much smaller than the rest of the stiffnesses ($k_i=1$ for $i>1$) for $\omega=1$. (a) Joint angles $\phi_i$, $i=1,2,\ldots,5$, versus time $t$. The dashed red line represents mean value $\bar{\phi}_1$ and in the inset we show zoom-in of the same plot (for joint angles $\phi_i$, $i=2,\ldots,5$), where the dashed green and maroon lines represent mean values $\bar{\phi}_2$ and $\bar{\phi}_3$, respectively. (b) The head angle $\theta$ versus $t$, showing oscillations about a constant-rate drift. The steady-state periodic solution is asymmetric about $\bar{\phi}_i=0$, leading to swimming on arc.} \label{fig15}
\end{figure*}

Finally, similarly to Purcell's swimmer we are interested in the existence of asymmetric solutions and bifurcations. For this reason, we consider an additional case, where in analogy to Purcell's swimmer we assume that the stiffness of the actuated joint is much smaller than those of the passive joints (but it does not vanish). Specifically, the input parameters in {\bf{case 9}} are:
\begin{equation}\label{E:case 10}
	\begin{aligned}
		&\tau_i=A_i \sin{(\omega t)}-k_i\phi_i,\\
		&A_1=1, \quad 
		A_2=A_3=A_4=A_5=0, \\
		&k_1=0.01, \quad k_2=k_3=k_4=k_5=1,
	\end{aligned}
\end{equation}
where we consider two frequencies $\omega=0.5$ and $\omega=1$, with the initial conditions  $\phi_1(0)=1$ [rad], $\phi_i(0)=0$, $i=2,\dots,5$.

In Fig.~\ref{fig14} we show the joint angles $\phi_i(t)$, $i=1,\dots,5$ versus time and $\theta(t)$ versus time for $\omega=0.5$. Again, it can be seen, that similarly to cases 1, 2, 7 and 8 (see Figs.~\ref{fig2}, \ref{fig3A}, \ref{fig10},  and~\ref{fig13}), after an initial transient the solution converges to symmetric oscillations about zero mean angles, $\bar{\phi}_i=0$, $i=1,\dots,5$, the head angle $\theta(t)$ oscillates about a constant value $\bar{\theta}$ indicating straight-forward net motion of the swimmer, and again as in case 8 the amplitudes of the joint angles decrease as we move from the head to the tail. However, if we increase the frequency to $\omega=1$ (with keeping the rest of parameters as in equation~\eqref{E:case 10}), we obtain a convergence to asymmetric periodic solution as can be seen in Fig.~\ref{fig15}, where we show the joint angles $\phi_i(t)$, $i=1,\dots,5$ versus time and the head angle $\theta(t)$ versus time for $\omega=1$. In particular, in Fig.~\ref{fig15} it can be observed that after an initial transient the joint angles $\phi_i(t)$, $i=1,2,3$ oscillate about nonzero mean values $\bar{\phi}_i\neq0$. Moreover, when $\omega=1$ the angle $\theta(t)$ does not oscillate about a constant value $\bar{\theta}$, but rather its mean per period is a decreasing function of time, indicating net motion of the swimmer along a curved arc (similarly to case 4 of Purcell's swimmer). Thus, Figs.~\ref{fig14} and~\ref{fig15} clearly indicate that in case 9 there exists a bifurcation in $\omega$ where for smaller values of $\omega$ the symmetric periodic solution is stable, and for larger values of $\omega$ the symmetric periodic solution becomes unstable and stable asymmetric periodic solutions emerge. This result was expected based on our results for Purcell's swimmer, and it emphasizes the fact that although a six-link swimmer is a much more complicated creature with much richer variety of dynamics, still our results for the simple three-link Purcell's swimmer provide us the intuition regarding some basic trends of expected behavior and bifurcations.

\section{Conclusions}\label{sec5}
In this study we first analyzed the dynamics of a microswimmer model with three rigid links (Purcell's swimmer), which is actuated by a periodic torque input in one or two joints, in parallel with torsional elasticity. Such mechanical inputs can also be interpreted as using proportional feedback in order to track periodic reference trajectories for the joint angles. We have used numerical integration of the swimmer's dynamics and numerical calculation of Poincar\'{e} maps in order to find steady-state periodic solutions and analyze their stability.

Under sinusoidal torque inputs, the swimmer's motion has crucial dependence on input's frequency, including frequencies at which the swimmer reverses its direction of motion as well as optimal frequencies for maximizing the mean swimming speed or net displacement per cycle, under symmetric oscillations about the straightened configuration. In addition, for cases where only one joint is active and its stiffness is much smaller than that of other passive joint, the system displays multiple periodic solutions, as well as stability transitions, bifurcations, and symmetry-breaking of periodic solutions upon varying the actuation frequency and relative stiffness at the joints. We also found cases where the swimmer's path curvature under asymmetric periodic solution reverses its sign upon varying the actuation frequency, which enables steering of the swimmer by modulating only the actuation frequency of a symmetric input. This phenomena, combined with the pitchfork bifurcation of symmetry breaking, may provide some possible simplified explanation to the way that flagellated bacteria can manipulate their periodic actuation in order to change their swimming direction and steering by exploiting buckling instability and symmetry breaking~\cite{Gadelha_2020,Son_2013}. This may also have relations to vibrational stabilization of flapping winged insects and micro-robotic fliers~\cite{Taha_2020}, and stability transitions in magnetically-actuated microswimmers~\cite{Harduf_2018,Paul_2023}, which are both based on parametric excitation.

Next, we extended our analysis to a six-link swimmer. For several sets of input parameters that were simulated we obtained a similar behavior to Purcell's swimmer. The similarity between the six-link swimmer and Purcell's swimmer includes reversing the direction of motion and existence of stable asymmetric solutions when the stiffness of the actuated joint is much smaller than those of the passive one, thus in particular indicating the existence of symmetry-braking bifurcation due to dynamic buckling instability.

Some challenges still remain open for future extensions of this research. First, experimental realization of the theoretical findings in a robotic swimmer with mechanical actuation and elasticity at the joints is an open challenging problem. Second, the numerical analysis calls for further exploration by applying asymptotic methods, aiming to obtain closed form approximate expressions that reflect the parametric influence on stability transitions and bifurcations, similar to the approach taken in~\cite{Harduf_2018,Paul_2023,Taha_2020,Tovi_2023}. In summary, it seems that Purcell's simple model and analysis based on the seminal work ``Life at low Reynolds number'' from 1977, is still alive and relevant nowadays and its results may be generalized to reflect interesting dynamic phenomena in swimming microorganisms as well as artificial micro-robots.

\section*{Acknowledgments}
This work was supported in part by Israel Science Foundation under Grant No. 1382/23, and Israel Ministry of Innovation, Science and Technology under Grant No. 3-17383.

\end{document}